\documentclass[sigconf]{acmart}

\usepackage{multirow}
\usepackage{array}
 
\usepackage{url}
\newcolumntype{M}[1]{>{\centering\arraybackslash}m{#1}} 
\newcolumntype{P}[1]{>{\raggedright\arraybackslash}p{#1}} 
\usepackage{xurl}

\usepackage{graphicx} 
\usepackage{booktabs} 
\usepackage{makecell}
\usepackage{ragged2e}
\usepackage{tabularx}
\usepackage{longtable}
\usepackage{float}
\usepackage{colortbl}  
\graphicspath{{./Figure/}} 

\AtBeginDocument{%
  }

\setcopyright{acmlicensed}

\copyrightyear{2026}
\acmYear{2026}
\setcopyright{cc}
\setcctype{by-nc-nd}
\acmConference[CHI EA '26]{Extended Abstracts of the 2026 CHI Conference on Human Factors in Computing Systems}{April 13--17, 2026}{Barcelona, Spain}
\acmBooktitle{Extended Abstracts of the 2026 CHI Conference on Human Factors in Computing Systems (CHI EA '26), April 13--17, 2026, Barcelona, Spain}
\acmDOI{10.1145/3772363.3798415}
\acmISBN{979-8-4007-2281-3/2026/04}
\begin{document}

\title{Analyzing the Presentation, Content, and Utilization of References in LLM-powered Conversational AI Systems}

\author{Jianheng Ouyang}
\email{jouyangae@connect.ust.hk}
\affiliation{%
  \institution{The Hong Kong University of Science and Technology}
  \country{Hong Kong S.A.R., China}
}

\author{Arpit Narechania}
\email{arpit@ust.hk}
\orcid{0000-0001-6980-3686}
\affiliation{%
  \institution{The Hong Kong University of Science and Technology}
  \country{Hong Kong S.A.R., China}
}

\renewcommand{\shortauthors}{Ouyang and Narechania}

\newcommand{\todo}[1]{\textcolor{red}{[Todo: #1]}}
\newcommand{\note}[1]{\textcolor{blue}{[Note: #1]}}
\newcommand{\que}[1]{\textcolor{red}{[Que: #1]}}
\newcommand{\arpit}[1]{\textcolor{brown}{[AN: #1]}}
\newcommand{\jianheng}[1]{\textcolor{blue}{[JO: #1]}}


\begin{abstract}
As conversational AI systems become popular for information retrieval and question-answering, the references they cite are key to ensuring their answers are reliable and trustworthy. Yet, no prior work systematically analyzes how these references are presented or their quality. We examine 1,517 references from 30 question–answer pairs across nine systems, focusing on their (1) presentation in the user interface and (2) quality using the CRAAP criteria. We find notable variations in the presentation, quality, and quantity of references across systems. For instance, ChatGPT provides more references (9.5 per response on average) with higher quality (15.48/20 CRAAP score), while Hunyuan-TurboS provides fewer references (4.0) and lower quality (11.65/20). Additionally, a preliminary user study shows that people rarely interact with these references and that their behavior differs across systems. These findings highlight the need for better interface designs that help users engage with and trust references more effectively.
\end{abstract}



\begin{teaserfigure}
  \centering
  \includegraphics[width=0.9\linewidth]{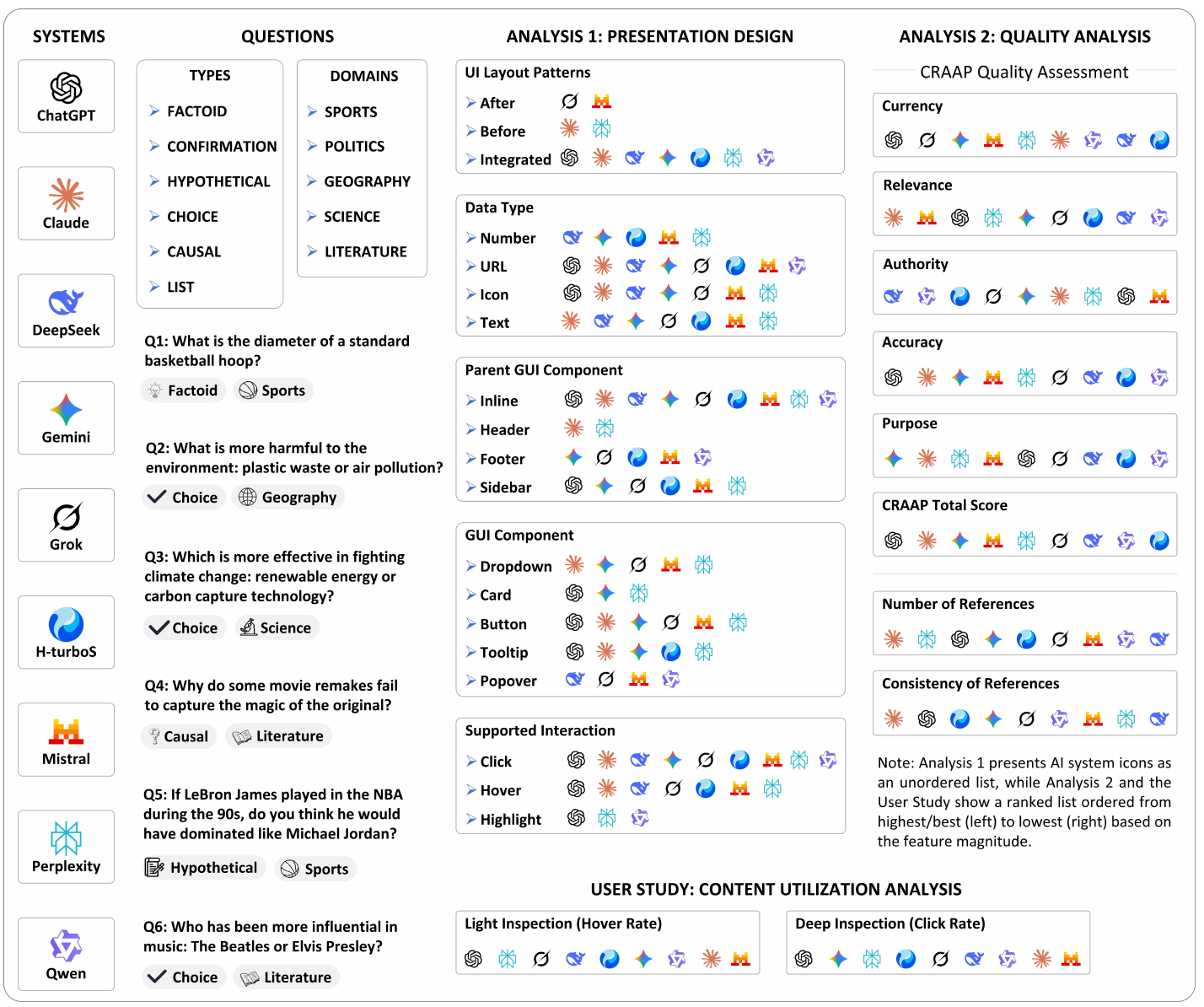}
  \caption{Summary of findings from three analyses of references across nine conversational AI systems. \textbf{Analysis 1: Presentation Design} illustrates how and where references are presented in the UI. \textbf{Analysis 2: Quality Analysis} evaluates the credibility of references using CRAAP criteria, alongside quantity and consistency metrics. \textbf{Content Utilization Analysis} reports results from a preliminary user study measuring participants' engagement with references. Together, these analyses reveal notable disparities in how references are presented, their reliability, and how users interact with them.}
  \label{fig:teaser}
\end{teaserfigure}

\maketitle

\section{Introduction and Background}
Search engines such as Google\footnote{\url{https://google.com}} and Bing\footnote{\url{https://bing.com}} have long supported information retrieval on the web~\cite{10.1145/1953122.1953146}. They present relevant hyperlinks, leaving users to find answers by themselves. In contrast, and more recently, conversational AI systems such as ChatGPT\footnote{https://chatgpt.com/} and Gemini\footnote{https://gemini.google.com/} facilitate question-answering by directly synthesizing answers, saving users significant effort in finding information manually~\cite{Vaswani2017Attention, brown2020language}. Today, billions of users worldwide rely on these systems~\cite{menlo2025consumer}: everyday users turn to them for general or specific information lookup~\cite{WOLF2024102821} and decision-making~\cite{huang2026webseek}; students use them to learn new concepts~\cite{YOUSSEF2024100316}; developers apply them for coding assistance~\cite{bakal2025experience}; and researchers employ them to find scientific literature~\cite{an2024vitality2}.
A key distinction between search engines and conversational AI systems lies in the role of references~\cite{metzger2024empowering}. In search engines, hyperlinks act as implicit references, delegating trust to the original websites. Conversely, conversational AI systems synthesize answers and select specific references to support them.
Because AI-generated answers can sometimes be incorrect or incomplete~\cite{10.1145/3770077,10.1145/3613904.3642459}, it becomes important for users to consult these references to verify the credibility of the answers and gain further knowledge~\cite{10.1145/3640543.3645200,li2025always}. Consequently, references play a more central role in shaping user trust and perceptions of credibility in conversational AI systems compared to traditional search engines~\cite{10.1609/aaai.v39i22.34550,huang-chang-2024-citation,desai2024cui,cai2022impacts}.

Notably, like the AI-generated answers, AI-generated references can also sometimes be flawed, i.e., they can be inaccurate, biased, irrelevant, fabricated, or poorly formatted~\cite{algaba2024large,walters2023fabrication,he2025getscitedgendermajoritybias,chen2023is}. For example, LLMs have been shown to exhibit gender bias by disproportionately referencing articles authored by male writers~\cite{he2025getscitedgendermajoritybias}. 
Beyond accuracy, other characteristics also matter. For example, reference quantity (number of references) can influence perceived thoroughness; consistency (the reliability of sources across repeated queries) can affect trust and reproducibility~\cite{liu-etal-2023-evaluating}. These issues highlight the need to systematically analyze how AI systems provide references. We therefore ask: \emph{``What is the quality of references provided by conversational AI systems?''}

Next, like the content of references, their presentation in the user interface (UI) is equally consequential. For example, thoughtful interaction design has been shown to reduce cognitive load and improve user comprehension, engagement, and trust in digital learning~\cite{priyadarshini2024impact,razaque5114814ai,yaqub2020effects,faudzi2023effects}. We aim to characterize the presentation and interaction for LLM-powered conversational AI systems. So we ask: \emph{``How do different conversational AI systems present references in the user interface for users to interact with?''}.

Lastly, while important, inspecting interface designs alone cannot reveal real-world behavior. Only through empirical observation can we understand how users actually perceive and utilize these references. Therefore, we ask: \emph{``How do users interact with references across different conversational AI systems?''}

Together, these three research questions span how references are generated, how they are surfaced in the interface, and how they are ultimately taken up (or ignored) by users, providing a holistic view of references in conversational AI systems.

To answer these questions, we conducted two system analyses and one content utilization user study across nine conversational AI systems~\cite{ThomasWolf2025}. We first examined the UI design choices associated with references, focusing on when they are included, their visual styles, placement, formatting, and interaction mechanisms. Next, we evaluated the quality of references using the CRAAP framework (Currency, Relevance, Authority, Accuracy, and Purpose)~\cite{blakeslee2004craap}, while also measuring the quantity and consistency of references provided by each system.
Lastly, we conducted a preliminary user study to understand how users interact with references across systems.

Our findings reveal notable differences across the nine systems, all summarized in Figure~\ref{fig:teaser}. ChatGPT and Claude achieved the highest overall quality, with CRAAP scores of 15.48/20, whereas Hunyuan-TurboS obtained the lowest score (11.65/20). Presentation strategies also varied, ranging from inline links to numbered footnotes and icons. Systems also exhibited divergent patterns in reference number and consistency: Claude maintained high and stable reference counts, whereas DeepSeek provided fewer references that fluctuated across repeated queries. Furthermore, our user study reveals that actual interaction--specifically hovering and clicking--remained generally low across all evaluated systems.

\begin{figure*}[ht]
  \centering
  \includegraphics[width=0.87\textwidth]{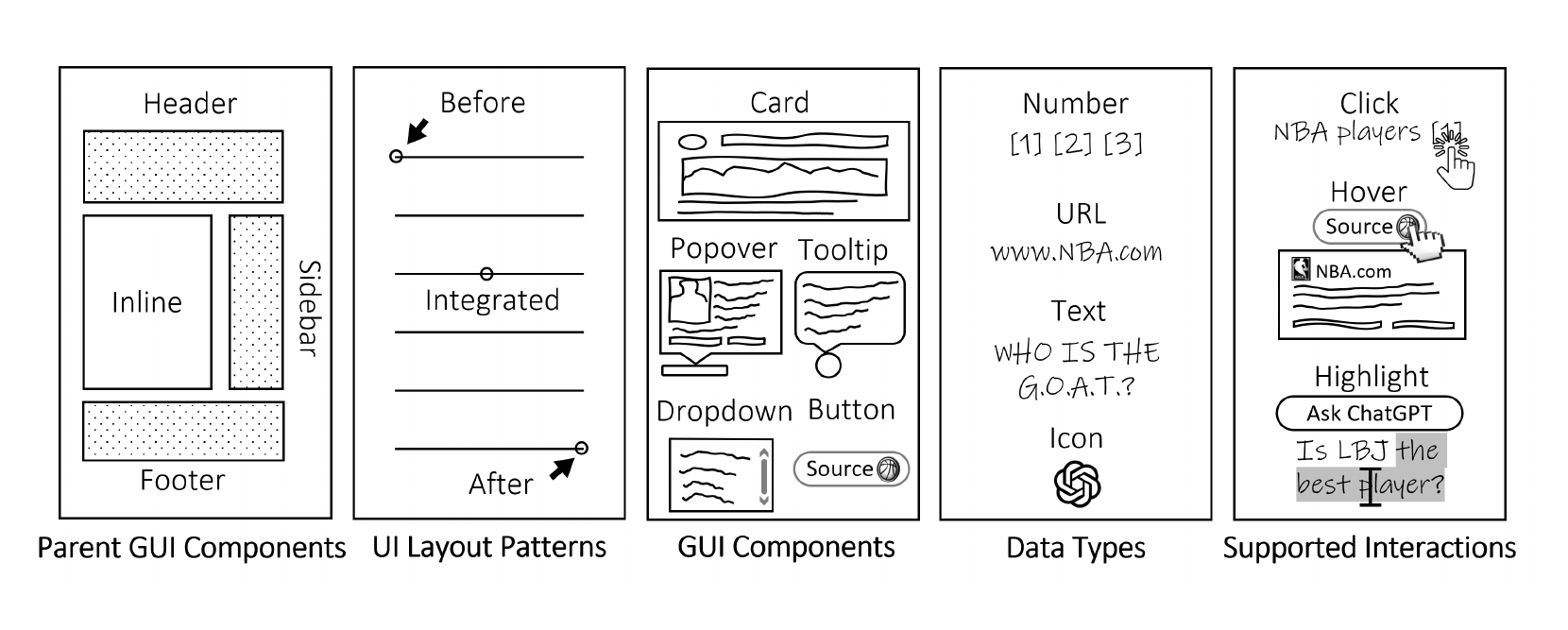}
  \caption{Five design dimensions for presenting references in conversational AI systems: (1) Parent GUI Components (header, inline, sidebar, footer), (2) Data Types (number, URL, text, icon), (3) UI Layout Patterns (before, integrated, after), (4) GUI Components (card, popover, tooltip, dropdown, button), and (5) Supported Interactions (click, hover, highlight). Each dimension represents a design decision that affects how users discover, access, and engage with references.}
  \label{fig:layout}
\end{figure*}

Collectively, these analyses provide insights into the quality and presentation of references in conversational AI systems. We posit that these differences may have implications for user trust and engagement with these systems~\cite{russell2024human}. Consequently, we call on the HCI community to study which UI layouts and interaction designs best support users, including the design of adaptive UIs that may tailor to specific or evolving user needs~\cite{narechania2024dissertation}. For example, systems can offer adaptive designs, such as inline references for students and concise answers for experts. Developers can also use CRAAP scores to ensure high-quality reference outputs. In doing so, we can collectively pave the way toward genuinely `user-centric AI'~\cite{narechania2025agentic}.
The primary contributions of this work include:
\begin{enumerate}
  \item Findings from an analysis about when and how LLM-powered conversational AI systems present references for users to view and interact with.
  \item Findings from an analysis about the quantity, consistency, and quality of the references output in LLM-powered conversational AI systems.
  \item Findings from a user study about the engagement levels and interaction patterns of users with references in LLM-powered conversational AI systems.
\end{enumerate}

\section{System Analyses}
\begin{figure*}
  \centering
  \includegraphics[width=0.6\textwidth]{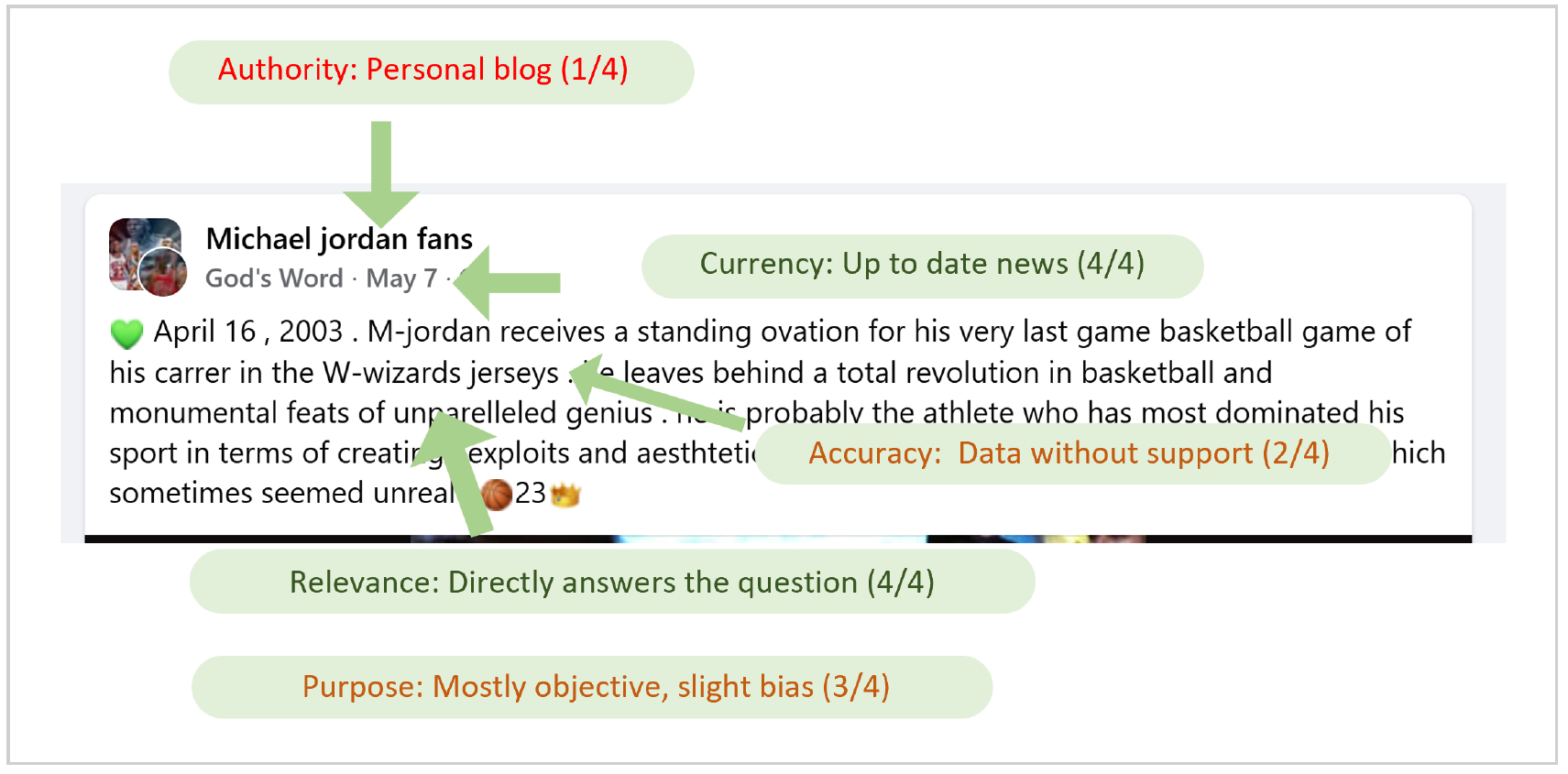}
    \caption{Quality assessment example using the CRAAP test. A social media post is evaluated across five criteria: \textbf{Authority} (1/4): a personal blog lacking formal credentials. \textbf{Currency} (4/4): up-to-date information. \textbf{Accuracy} (2/4): claims lack sufficient supporting evidence. \textbf{Relevance} (4/4): directly addresses the subject. \textbf{Purpose} (3/4): largely objective but shows slight bias.}
    \label{fig:craapexample}
\end{figure*}
\subsection{Methodology}

First, we selected nine conversational AI systems based on market visibility and user adoption~\cite{ThomasWolf2025}: \textit{ChatGPT-4o, Claude 3.5 Sonnet, DeepSeek-V3, Gemini 2.5, Grok 3, Hunyuan-TurboS, Mistral Medium3, Perplexity AI}, and \textit{Qwen3-235B-A22B-2507}. 
Next, to study AI systems' outputs on different types of questions, we curated a query set covering two dimensions: (1) six question types based on the taxonomy proposed by Mohasseb~et~al.~\cite{MOHASSEB20181228}--\textit{factoid, confirmation, hypothetical, choice, causal, and list} and (2) five domains--\textit{sports, politics, geography, science, and literature} (Figure~\ref{fig:teaser}). 
Subsequently, we prompted each of the 9 systems once per query, extracting references and metadata through direct UI interaction\footnote{All data was collected between July 1-7, 2025.}. This initial round yielded 270 system responses ($30 \text{ questions} \times 9 \text{ systems}$), containing 1,517 unique references used for our primary UI and quality analyses. Lastly, to evaluate reference consistency, we used independent-session repeats: for each question, we opened a fresh chat, asked the query, and recorded the number of references. In total, we conducted 1,350 trials ($30 \text{ questions} \times 9 \text{ systems} \times 5 \text{ iterations}$) to examine how reference counts fluctuate across repeated conversations. We evaluated the internal consistency of these reference counts across iterations for each system using Cronbach's $\alpha$~\cite{cronbach1951coefficient}.

Following extraction, one author analyzed the data to perform two complementary analyses about: (1) \textit{Presentation Design} to identify patterns in layout, interactivity, and data type; and (2) \textit{Content Analysis} to evaluate reference credibility and quality. 
For the former, one author visually inspected the UI screenshots for each system response, and developed a codebook of reference attributes--such as UI layout and data types--which was finalized in consultation with a second author, resolving any discrepancies via consensus.
Similarly, for the latter, one author employed the CRAAP framework~\cite{blakeslee2004craap}--a widely-used standard for evaluating source credibility--assessing each of its five dimensions on a 4-point scale (1 to 4, as illustrated in Figure~\ref{fig:craapexample}): Currency (timeliness), Relevance (topic alignment), Authority (source expertise), Accuracy (factual correctness), and Purpose (objectivity)--finalized via consensus with the second author.
Although CRAAP is designed for evaluating human-authored sources that actually exist~\cite{blakeslee2004craap}, we adapted it for AI-generated references that include hallucinated or inaccessible references (e.g., 404 errors). We assigned such references the minimum score (1/5) for Accuracy and Relevance--the dimensions directly undermined by non-existence--while Currency, Authority, and Purpose were assessed from available metadata. We verified existence by accessing each URL and cross-checking details against known publications.

\subsection{Findings: Presentation Design}

We identified five design dimensions to characterize how systems present references. Figure~\ref{fig:layout} summarizes them, and Figure~\ref{fig:teaser} shows how systems align with each. Each dimension is described below.

\begin{itemize}
    \item \textbf{Parent GUI Components}: This dimension describes the placement of references within the interface. All nine systems used \textit{Inline} references within the text. To provide additional details without clutter, several systems (e.g., Grok, Gemini) also employed \textit{Sidebar} (66.7\%, $N=6$) or \textit{Footer} (55.6\%, $N=5$) placements. Only two systems positioned references in a \textit{Header} (22.2\%, $N=2$).
    
    \item \textbf{UI Layout Patterns}: This dimension characterizes the arrangement of references relative to the main content. Most systems (77.8\%, $N=7$), including ChatGPT, adopted an \textit{Integrated} layout where references are embedded within responses. Two systems placed references \textit{Before} the answer (22.2\%, $N=2$), as observed in Claude, while another two positioned them \textit{After} the answer (22.2\%, $N=2$), as in Grok.
    
    \item \textbf{GUI Components}: This dimension concerns the interactive elements used to display references. Most systems implemented \textit{Button} components (66.7\%, $N=6$) to facilitate interaction with references. \textit{Dropdown} (55.6\%, $N=5$) and \textit{Tooltip} (55.6\%, $N=5$) were also common for revealing additional details, while \textit{Popover} (44.4\%, $N=4$) was used by some systems (e.g., Grok, Qwen) for richer previews. Only three systems employed \textit{Card} components (33.3\%, $N=3$).
    
    \item \textbf{Data Types}: This dimension identifies the kinds of information shown for each reference. Most systems, including ChatGPT and Gemini, prioritized clickable \textit{URL} links (88.9\%, $N=8$) for direct access to references. \textit{Icon} (77.8\%, $N=7$) and \textit{Text} (77.8\%, $N=7$) representations were also prevalent to balance recognition and description. Five systems (e.g., DeepSeek) used \textit{Number} indicators (55.6\%, $N=5$) as compact inline markers.
    
    \item \textbf{Supported Interactions}: This dimension specifies the user interactions supported for referencing. All systems supported basic \textit{Click} interactions for navigation. Most (77.8\%, $N=7$) also supported \textit{Hover} actions for previews. Only three systems (33.3\%, $N=3$), such as Qwen, enabled \textit{Highlight} interactions for operations like citing or quoting selected text.
\end{itemize}

These findings reveal diverse design strategies across the nine systems, reflecting distinct approaches to balancing reference visibility with reading flow. Some prioritize integrated inline references that enhance discoverability but risk cluttering the main content, while others favor sidebar or footer placements that preserve response readability at the cost of reduced prominence. Ultimately, no single strategy is optimal, highlighting the need for adaptive designs that dynamically adjust to user and task demands.
\begin{figure*}
  \centering
  \includegraphics[width=\textwidth]{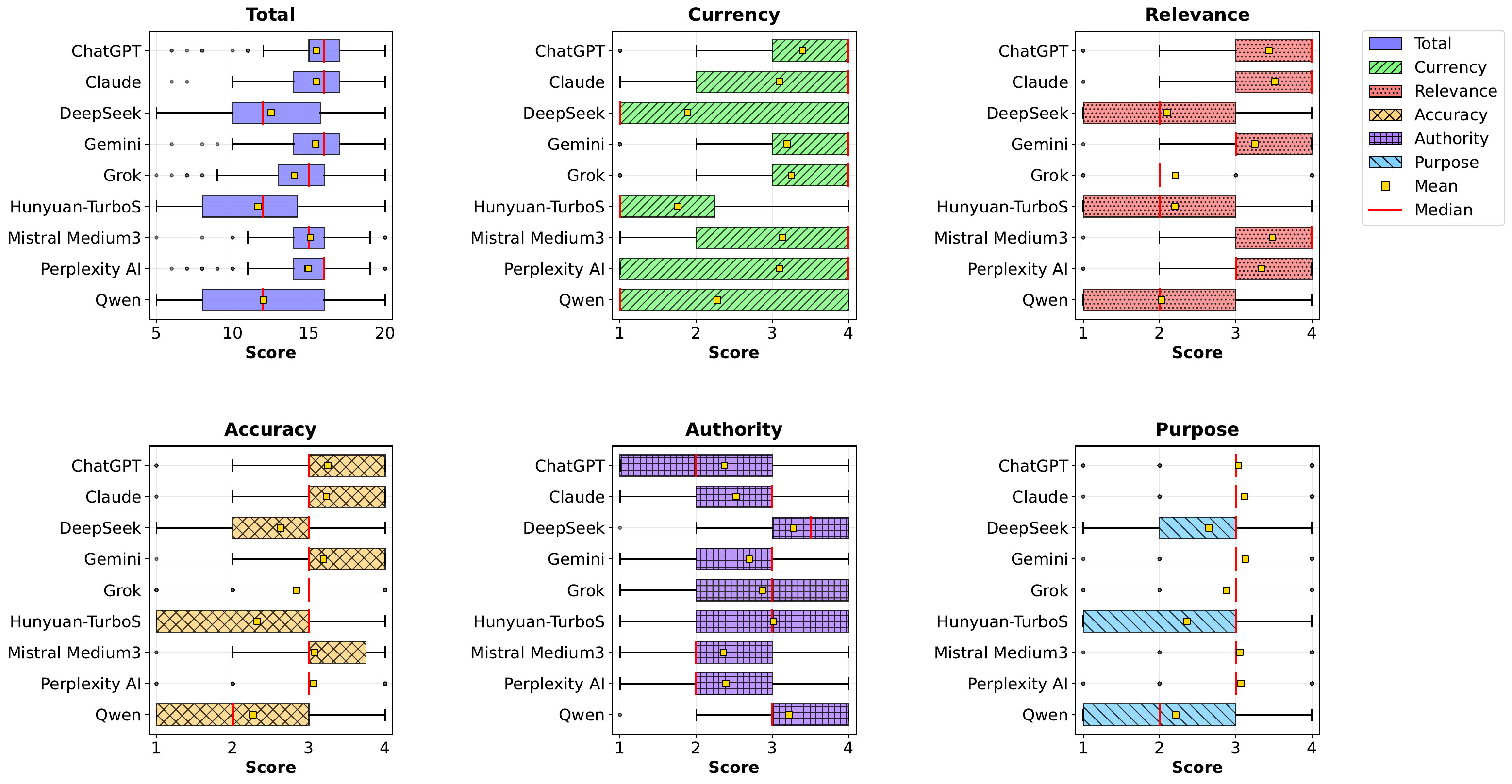}
  \caption{Quality of references from nine conversational AI systems across the CRAAP Framework. This figure presents six horizontal box plots, visualizing the score distribution for each conversational AI system on the CRAAP criteria (Currency, Relevance, Authority, Accuracy, Purpose) and a Total score. Each box plot shows the median (red line), mean (yellow square), interquartile range (box), and outliers (black dots). For metrics where the box appears as a single line, it indicates that the 25th and 75th percentiles coincide, meaning a high consistency in scores for the middle 50\% of observations. Observations show that Western models generally outperform others, with ChatGPT and Claude showing high consistency across all criteria.}
  \label{fig:raincloud}
\end{figure*}
\subsection{Findings: Quality Analysis}
\label{sec:quality-analysis}

We assessed the reference outputs of the nine selected systems using the CRAAP framework (Currency, Relevance, Authority, Accuracy, and Purpose). We found notable differences in reference quality across systems, summarized in Figure~\ref{fig:raincloud}, and described below:

\begin{itemize}
    \item \textbf{Currency}: ChatGPT scored highest ($3.39 \pm 1.10$), followed by Grok ($3.25 \pm 1.17$) and Gemini ($3.19 \pm 1.25$). Hunyuan-TurboS had the lowest score ($1.76 \pm 1.29$), indicating inconsistency in the timeliness of references.
    \item \textbf{Relevance}: Claude achieved the highest relevance score ($3.52 \pm 0.60$), followed by Mistral Medium3 ($3.48 \pm 0.67$) and ChatGPT ($3.43 \pm 0.68$). Qwen had the lowest score ($2.03 \pm 1.05$), suggesting weaker alignment between content and question.
    \item \textbf{Authority}: DeepSeek obtained the highest authority score ($3.27 \pm 0.83$), followed by Qwen ($3.22 \pm 0.80$) and Hunyuan-TurboS ($3.02 \pm 0.97$). Mistral Medium3 scored lowest ($2.36 \pm 1.03$), reflecting reliance on less credible sources.
    \item \textbf{Accuracy}: ChatGPT achieved the highest accuracy ($3.25 \pm 0.69$), followed by Claude ($3.23 \pm 0.55$) and Gemini ($3.19 \pm 0.63$). Qwen had the lowest accuracy ($2.27 \pm 1.18$), indicating a higher incidence of factual errors.
    \item \textbf{Purpose}: Claude and Gemini scored highest ($3.12 \pm 0.48$ and $3.12 \pm 0.55$), followed by Perplexity ($3.07 \pm 0.59$). Qwen had the lowest score ($2.21 \pm 1.14$), indicating less objective information presentation.
\end{itemize}


In addition to the CRAAP \textbf{quality} dimensions, we also evaluated the \textbf{quantity} and \textbf{consistency} of references. Quantity captures the average number of citations a system produces per response, while consistency reflects the stability of that count across multiple iterations. Consistency was assessed using \textit{Cronbach's} $\alpha$, a reliability coefficient ranging from $0$ to $1$ that indicates the internal stability of repeated outcomes~\cite{adamson2013reliability,schroeder2025trustllmjudgmentsreliability}. Values above $0.70$ are generally considered acceptable.

We found that Claude produced exactly 10 references in every iteration, demonstrating perfect mathematical consistency ($\alpha = 1.000$). However, this lack of variation likely reflects a rigid, hardcoded constraint rather than adaptable behavior. In comparison, ChatGPT generated an average of 9.5 references with excellent consistency ($\alpha = 0.961$), suggesting stable yet dynamic reference generation. Interestingly, while Hunyuan-TurboS provided notably fewer references (4.0 on average), it maintained a similarly high level of consistency ($\alpha = 0.960$). Conversely, DeepSeek provided the fewest references (2.1 on average) and exhibited the lowest consistency ($\alpha = 0.811$), indicating significant fluctuations across runs.
Overall, there were notable differences in terms of quality, quantity, and consistency of referenecs across systems. ChatGPT emerged as the most robust performer, offering high CRAAP scores alongside stable and numerous reference counts. Claude also performed highly but exhibited some rigidity in the quantity of references. In contrast, DeepSeek and Qwen produced fewer, lower-quality references with greater variability.

\section{Content Utilization Analysis}

Having analyzed how systems present references in the UI and their actual quality, we next investigated how users interact with these references during a series of information-seeking tasks.
\begin{figure*}
  \centering
  \includegraphics[width=0.6\textwidth]{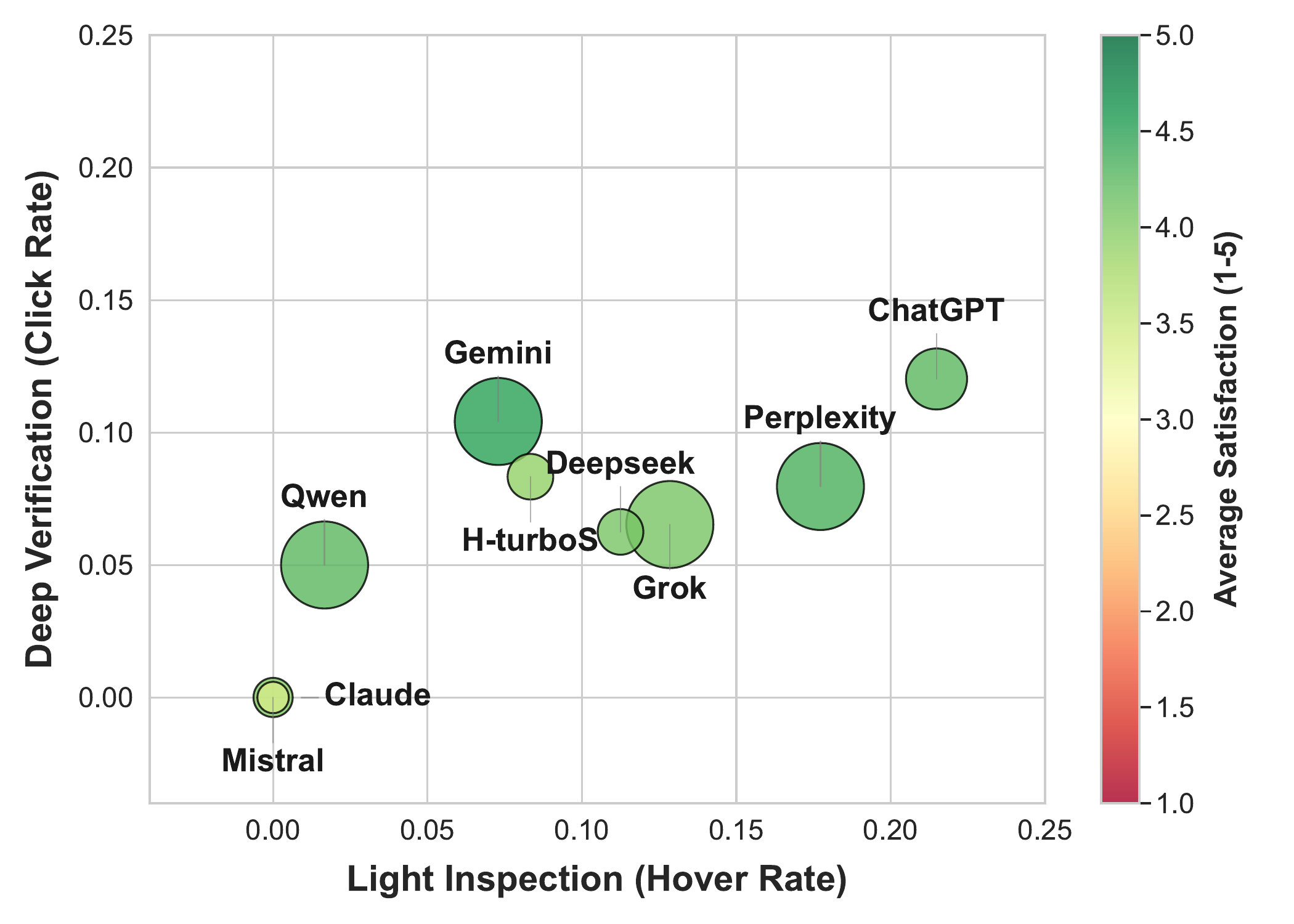}
  \caption{User Interaction Behavior. The bubble chart shows Light Inspection (Hover Rate) against Deep Verification (Click Rate) across nine AI systems. Bubble size represents reference counts, while color indicates average user satisfaction (1-5).}
  \label{fig:bubble_chart}
\end{figure*}
\subsection{Methodology}
We recruited 12 undergraduate students (8 male, 4 female) via university email. 
Participants were asked to use each of the nine AI systems to answer a curated set of questions spanning across multiple domains and question types~(as illustrated in Figure~\ref{fig:teaser}). 
For example, participants were asked specific questions such as, \textit{``Why do leaves change color in the fall across different tree species and regions?''} and \textit{``Which is more effective in fighting climate change: renewable energy or carbon capture technology?''} Note that the order of questions was randomized to mitigate presentation bias.

For each question-answer pair, we instructed participants to study the latter and evaluate its accuracy. We provided no instructions about the presence or position of references; we also did not ask participants to ask follow-up questions to fetch missing references, ensuring that all engagement--including the decision to forgo verification--was natural and entirely self-directed. To help us analyze user behavior, we recorded hover and click frequencies (proxies for visual attention and active verification), reference counts, and 5-point Likert satisfaction scores. 
Detailed questions and findings are available in supplemental material.

\subsection{Findings}

Figure~\ref{fig:bubble_chart} summarizes the user study results. Verification rates were consistently low across systems, with hover and click frequencies remaining below 25\% for all platforms. ChatGPT achieved the highest engagement ($\sim$22\% hover, $\sim$12\% click). Hover interactions consistently exceeded clicks, suggesting a stronger preference for quick tooltips or popovers over direct source access~\cite{10.1145/1571941.1571951,he2025transparencyequalsourcepresentation}.

Despite limited verification activity, satisfaction scores remained high (mostly $>$4.0; Figure~\ref{fig:bubble_chart}), revealing a potential \textit{trust gap} in which users accepted AI-generated content without validation~\cite{10.1016/j.chb.2024.108352,passi2022overreliance,narechania2025guidancesourcematters}. Participants often agreed with the system's answer without viewing references to confirm its claims. This trend persisted even for systems with lower reference accuracy (e.g., Qwen; Section~\ref{sec:quality-analysis}), indicating that plausibility often outweighed verification.

Overall, these results suggest that users relied more on perceived plausibility than on explicit source checking~\cite{doi:10.1073/pnas.2518443122,fan2026aipersuadesadversarialexplanation,10.1145/3449287}. This behavior reflects a tendency to treat conversational AIs as authoritative experts rather than informational aids, risking over-reliance~\cite{10.1145/3733567.3735566,10.1145/3706598.3714082}--though this requires further study.

\section{Limitations}

Our content analysis and user study had many limitations; first, given the rapid evolution of conversational AI systems, our findings represent a snapshot of the landscape as of July 2025; future model updates may cause specific reference behaviors to shift~\cite{chen2023chatgptsbehaviorchangingtime}. Next, while our methodology penalized hallucinated references with the lowest \textit{Accuracy} and \textit{Relevance} scores, we recognize that conversational AI systems inherently produce fabricated content. Traditional frameworks like CRAAP are designed with the assumption that the evaluated sources actually exist. Therefore, because relying on frameworks built for human-authored content is not a long-term solution, we emphasize the critical need to establish specialized evaluation metrics and rules specifically designed for references provided by AI systems.

Next, although our investigation relies on a fixed and focused set of questions, tools, and participants, our findings provide preliminary evidence that the disparity between reference presentation and quality is a critical issue and establishing this as an important area for future large-scale investigation. Additionally, interaction behavior is currently modeled solely based on mouse cursor inputs, which may not be a complete proxy for attention~\cite{10.1145/1978942.1979125}. Future work may consider integrating eye-tracking technology to more accurately characterize user attention~\cite{10.1109/TVCG.2021.3114827}. Future research should also explore how specific UI interventions, such as credibility warning labels~\cite{10.1145/3613904.3642473} or interactive visualization of source provenance~\cite{liao2023ai}, can encourage users toward more active verification behaviors.

\section{Conclusion}

We analyzed 1,517 references across nine LLM-based conversational AI systems to characterize the presentation, quality, quantity, and consistency of references in the user interface. We found notable differences between systems across all metrics; however, overall, current designs often relegate references to static decorations rather than first-class interactive elements necessary for active verification. These results advocate for adaptive UIs~\cite{narechania2024dissertation} that dynamically tailor reference granularity to user needs and expertise.

\bibliographystyle{ACM-Reference-Format}
\bibliography{references}

\end{document}